\documentclass[prl,aps,twocolumn]{revtex4}

\usepackage{bm}
\usepackage{amsmath}
\usepackage{amssymb}

\usepackage{graphicx}
\begin{document}



\title{Pairing symmetry of superconductivity coexisting with antiferromagnetism}


\author{Keisuke Shigeta}
\author{Seiichiro Onari}
\author{Yukio Tanaka}

\affiliation{Department of Applied Physics, 
Nagoya University, Nagoya 464-8603, Japan}

\begin{abstract}
Pairing symmetry in the superconducting state 
coexisting with antiferromagnetic order is studied  
based on a microscopic theory. 
We calculate the linearized $\mathrm{\acute{E}}$liashberg's equation 
within the random phase approximation in the Hubbard model 
with a staggered field. 
We find that odd-frequency spin-triplet (equal-spin) $s$-wave pairing state 
can be realized. 
This result contradicts a naive expectation 
that antiferromagnetic order induces 
antiferromagnetic spin fluctuation and favors spin-singlet $d$-wave pairing as 
in the standard strongly correlated systems.  
\end{abstract}

\maketitle
%
%
In condensed matter physics, coexistence of superconductivity (SC) and 
magnetism has been a fundamental issue attracting a great deal of attention 
\cite{Ginzburg}. 
While SC coexists with ferromagnetism in, $e.g.$, UGe$_2$ and URhGe 
\cite{UGe2,URhGe}, 
SC coexists with antiferromagnetism (AF) in, $e.g.$, CeCu$_2$Si$_2$ and 
CeRhIn$_5$ \cite{CeCu2Si2,CeRhIn5}. 
Considering the fact that antiferromagnetic spin fluctuation mediates 
spin-singlet pairing, we simply expect that coexistence of antiferromagnetic 
order (AFO) makes spin-singlet $d$-wave pairing more stable. 
However, pairing symmetry of Cooper pair in SC coexisting with AF 
is not simple. 
In CeCu$_2$Si$_2$ and 
CeRhIn$_5$ \cite{CeCu2Si2_NQR,CeRhIn5_NQR,CeRhIn5_specific_heat}, 
gapless excitation of quasiparticle has been reported experimentally. 
It has been pointed theoretically that 
a competition between even-frequency (even-$\omega$) spin-singlet $d$-wave 
and odd-frequency (odd-$\omega$) \cite{Berezinskii} spin-singlet $p$-wave 
pairings is possible \cite{Fuseya} in the superconducting state 
due to the coexistence of AF. 
\par
Odd-$\omega$ pairing has been originally proposed by Berezinskii in 1974 
\cite{Berezinskii}. 
A possibility of odd-$\omega$ spin-triplet pairing has been discussed 
in the context of superfluid $^3$He \cite{Berezinskii}. 
After that, Balatsky and Abrahams have proposed odd-$\omega$ spin-singlet 
pairing \cite{odd1}. 
Starting from these proposals, a lot of intensive studies about 
odd-$\omega$ pairing have been done both in bulk \cite{Fuseya,odd2,odd3,odd4,Hotta,Solenov,Kusunose} and inhomogeneous \cite{odd5,odd6,odd7} systems. 
\par
Although there have been several studies about SC coexisting with 
AFO, pairing symmetry of Cooper pair 
has not been fully resolved based on a microscopic theory. 
In the present paper, we study pairing symmetry in the superconducting state 
coexisting with AF focusing on the possible odd-$\omega$ spin-triplet pairing 
based on a microscopic theory. 
We consider the Hubbard model with a staggered field originating from 
commensurate AFO 
and solve the linearized $\mathrm{\acute{E}}$liashberg's 
equation within the random phase approximation (RPA). 
We find that a competition between even-$\omega$ 
spin-singlet $d$-wave and odd-$\omega$ spin-triplet (equal-spin) $s$-wave 
pairings is induced by the staggered field. 
We show that odd-$\omega$ equal-spin-triplet 
$s$-wave pairing state becomes the most dominant 
under a high staggered field due to the charge fluctuation induced by AFO.
This result contradicts a naive expectation that AFO 
induces antiferromagnetic spin fluctuation and favors spin-singlet $d$-wave 
pairing. 
\par
\begin{figure}[htbp]
\includegraphics[width=0.99\linewidth,keepaspectratio]
                  {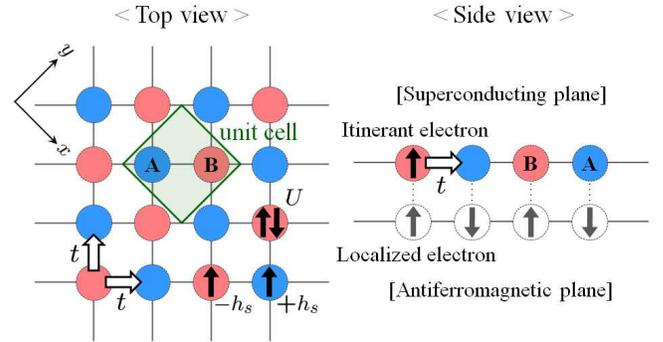}
 \caption{(Color online) The model considered in the present study with a hopping $t$, the on-site Coulomb repulsion $U$, and a staggered field $h_s$.}
\label{fig_lattice}
\end{figure}

%
%
We model a system where SC coexists with AF by separately considering 
superconducting and antiferromagnetic electrons, $i.e.$, 
we assume that interactions between superconducting and antiferromagnetic 
electrons are not so strong. 
We apply the Hubbard model to 
itinerant electrons generating superconductivity 
taking into account a background staggered field 
due to AFO by localized electrons as shown in Fig. \ref{fig_lattice}. 
We consider the nearest neighbor hopping $t$ and the on-site Coulomb repulsion $U$ and assume that the 
staggered field by AFO is a given quantity $h_s$ without solving dynamics of 
localized electrons. 
There are two kinds of sublattices A and B under finite $h_s$. 
Potential felt by 
an itinerant electron with spin $\sigma$ is shifted by the staggered field 
$+(-)s_{\sigma} h_s$ on a sublattice A (B) with 
$s_{\uparrow(\downarrow)}=+(-)1$, $i.e.$, a sublattice A is 
$\downarrow$-electron-rich while a sublattice B is 
$\uparrow$-electron-rich under finite $h_s$. 
The corresponding Hamiltonian is given by 
\begin{align}
{\cal{H}}=&-t\sum_{\langle i,j\rangle,\sigma}(c_{i\sigma}^{\dagger}c_{j\sigma}+{\mathrm{H.c.}})+U\sum_{i}n_{i\uparrow}n_{i\downarrow} \notag \\
&+h_s\sum_{i,\sigma}s_{\sigma} \tau_i c_{i\sigma}^{\dagger}c_{i\sigma}, 
\end{align}
where $c_{i\sigma}^{(\dagger )}$ annihilates (creates) an itinerant 
electron with spin 
$\sigma$ at a site $i$ and $n_{i\sigma}$ is a number operator. 
$\tau_i=1$ on a sublattice A while $\tau_i=-1$ on a sublattice B. 
The Green's function without $U$ is given by
\begin{eqnarray}
\hat{G}_{\sigma}(k)&=&\left[ {\mathrm{i}}\omega_n\hat{I}-\hat{\varepsilon}_{{\boldsymbol{k}}\sigma}+\mu\hat{I}\right]^{-1}, 
\label{eq_green}
\\
\varepsilon_{{\boldsymbol{k}}\sigma}^{\mathrm{AA}}&=&s_{\sigma} h_s, \\
\varepsilon_{{\boldsymbol{k}}\sigma}^{\mathrm{AB}}&=&-t-t\exp(-{\mathrm{i}}k_xa)-t\exp(-{\mathrm{i}}k_ya) \notag \\
&&-\mspace{4mu}t\exp(-{\mathrm{i}}k_xa-{\mathrm{i}}k_ya), \\
\varepsilon_{{\boldsymbol{k}}\sigma}^{\mathrm{BA}}&=&-t-t\exp({\mathrm{i}}k_xa)-t\exp({\mathrm{i}}k_ya) \notag \\
&&-\mspace{4mu}t\exp({\mathrm{i}}k_xa+{\mathrm{i}}k_ya), \\
\varepsilon_{{\boldsymbol{k}}\sigma}^{\mathrm{BB}}&=&-s_{\sigma} h_s,
\end{eqnarray}
where $k\equiv ({\mathrm{i}}\omega_n,{\boldsymbol{k}})$ is 
an ellipsis notation with the fermionic Matsubara frequency 
$\omega_n=(2n-1)\pi T$ and the momentum ${\boldsymbol{k}}$, 
$\mu$ is the chemical potential, 
$\varepsilon_{{\boldsymbol{k}}\sigma}^{\alpha\beta}$ is 
an $(\alpha,\beta)$-element of $\hat{\varepsilon}_{{\boldsymbol{k}}\sigma}$, 
$\hat{I}$ is a unit matrix, and $a$ is the lattice constant of the unit cell. 
\par
Within the RPA, we solve the linearized $\mathrm{\acute{E}}$liashberg's 
equation
\begin{align}
&\lambda\Delta_{\sigma\sigma'}^{\alpha\beta}(k) \notag \\
&\mspace{10mu}=-\frac{T}{N}\sum_{k,\alpha',\beta'}V_{\sigma\sigma'}^{\alpha\beta}(k,k')G_{\sigma}^{\alpha\alpha'}(k')G_{\sigma'}^{\beta\beta'}(-k')\Delta_{\sigma\sigma'}^{\alpha'\beta'}(k'),
\label{eq_eliash}
\end{align}
where $\lambda$ is an eigenvalue and $N$ is the number of sites. 
$\Delta_{\sigma\sigma'}^{\alpha\beta}(k)$ 
and $V_{\sigma\sigma'}^{\alpha\beta}(k,k')$ 
are $(\alpha,\beta)$-elements of a gap function 
$\hat{\Delta}_{\sigma\sigma'}(k)$ and an effective pairing interaction 
$\hat{V}_{\sigma\sigma'}(k,k')$, respectively.  
$T$ is equal to the superconducting transition temperature 
$T_{\mathrm{C}}$ when $\lambda$ reaches unity. 
Thus, larger $\lambda$ corresponds to more stable superconducting state. 
In Eq. (\ref{eq_eliash}), a Cooper pair with equal (opposite) spins 
requires $\sigma'=\sigma(\bar{\sigma})$. 
$\hat{V}_{\sigma\sigma'}(k,k')$ is given by
\begin{align}
\hat{V}_{\sigma\bar{\sigma}}(k,k')=&\hat{U}-U^2\hat{\chi}_{\bar{\sigma}\sigma}^{\mathrm{bub}}(k-k')-U^2\hat{\chi}_{\sigma\bar{\sigma}}^{\mathrm{lad}}(k+k'), 
\label{eq_interaction_ap}
\\
\hat{V}_{\sigma\sigma}(k,k')=&-U^2\hat{\chi}_{\bar{\sigma}\bar{\sigma}}^{\mathrm{bub}}(k-k'),
\label{eq_interaction_p}
\end{align}
where $\hat{U}\equiv U\hat{I}$ and $\hat{\chi}_{\sigma\sigma'}^{\mathrm{bub}}(q)$ ($\hat{\chi}_{\sigma\bar{\sigma}}^{\mathrm{lad}}(q)$) 
denotes a part of susceptibility 
from bubble-type (ladder-type) diagrams. 
Here, $q\equiv ({\mathrm{i}}\nu_m,{\boldsymbol{q}})$ is 
an ellipsis notation with the bosonic Matsubara frequency 
$\nu_m=2m\pi T$ and the momentum ${\boldsymbol{q}}$. 
$\hat{\chi}_{\sigma\sigma'}^{\mathrm{bub}}(q)$ and $\hat{\chi}_{\sigma\bar{\sigma}}^{\mathrm{lad}}(q)$ are given by
\begin{align}
\hat{\chi}_{\sigma\sigma}^{\mathrm{bub}}(q)=&\hat{\chi}_{\sigma\sigma}^0(q)\left[\hat{I}-\hat{U}\hat{\chi}_{\bar{\sigma}\bar{\sigma}}^0(q)\hat{U}\hat{\chi}_{\sigma\sigma}^0(q)\right]^{-1}, \\
\hat{\chi}_{\sigma\bar{\sigma}}^{\mathrm{bub}}(q)=&-\hat{\chi}_{\sigma\sigma}^{\mathrm{bub}}(q)\hat{U}\hat{\chi}_{\bar{\sigma}\bar{\sigma}}^0(q), \\
\hat{\chi}_{\sigma\bar{\sigma}}^{\mathrm{lad}}(q)=&-\hat{\chi}_{\bar{\sigma}\sigma}^0(q)\left[\hat{I}-\hat{U}\hat{\chi}_{\bar{\sigma}\sigma}^0(q)\right]^{-1},
\end{align}
where $\hat{\chi}_{\sigma\sigma'}^0(q)$ is the irreducible susceptibility. 
An $(\alpha,\beta)$-element of $\hat{\chi}_{\sigma\sigma'}^0(q)$ is given by
\begin{align}
\chi_{\sigma\sigma'}^{0,\alpha\beta}(q)=-\frac{T}{N}\sum_{k}G_{\sigma}^{\alpha\beta}(k+q)G_{\sigma'}^{\beta\alpha}(k).
\end{align}
$\hat{\chi}_{\sigma\sigma'}^{\mathrm{bub}}(q)$ and $\hat{\chi}_{\sigma\bar{\sigma}}^{\mathrm{lad}}(q)$ give the transverse spin, longitudinal spin, and charge susceptibilities written as
\begin{align}
\hat{\chi}_{\mathrm{sp}}^{+-}(q)=&-\hat{\chi}_{\sigma\bar{\sigma}}^{\mathrm{lad}}(q), 
\label{eq_chi_sp_+-}
\\
\hat{\chi}_{\mathrm{sp}}^{zz}(q)=&\frac{1}{2}\left[\hat{\chi}_{\uparrow\uparrow}^{\mathrm{bub}}(q)+\hat{\chi}_{\downarrow\downarrow}^{\mathrm{bub}}(q)-\hat{\chi}_{\uparrow\downarrow}^{\mathrm{bub}}(q)-\hat{\chi}_{\downarrow\uparrow}^{\mathrm{bub}}(q)\right], 
\label{eq_chi_sp_zz}
\\
\hat{\chi}_{\mathrm{ch}}(q)=&\frac{1}{2}\left[\hat{\chi}_{\uparrow\uparrow}^{\mathrm{bub}}(q)+\hat{\chi}_{\downarrow\downarrow}^{\mathrm{bub}}(q)+\hat{\chi}_{\uparrow\downarrow}^{\mathrm{bub}}(q)+\hat{\chi}_{\downarrow\uparrow}^{\mathrm{bub}}(q)\right].
\label{eq_chi_ch}
\end{align}
Note that the antiferromagnetic moment is assumed to be parallel to 
the spin quantization axis ($z$ axis) in the present study. 
$\hat{\chi}_{\mathrm{sp}}^{+-}(q)$ and $\hat{\chi}_{\mathrm{sp}}^{zz}(q)$ 
indicate spin fluctuations in a $xy$ plane and a $z$ direction while 
$\hat{\chi}_{\mathrm{ch}}(q)$ indicates charge fluctuation. 
We define the maximum eigenvalue of 
$\hat{U}\hat{\chi}_{\bar{\sigma}\bar{\sigma}}^0(q)\hat{U}\hat{\chi}_{\sigma\sigma}^0(q)$ 
($\hat{U}\hat{\chi}_{\bar{\sigma}\sigma}^0(q)$) 
as the Stoner's factor $f_{\mathrm{S}}^{\mathrm{bub}}$ 
for $\hat{\chi}_{\sigma\sigma'}^{\mathrm{bub}}(q)$ 
($f_{\mathrm{S}}^{\mathrm{lad}}$ for $\hat{\chi}_{\sigma\bar{\sigma}}^{\mathrm{lad}}(q)$) 
and the larger one as the Stoner's factor $f_{\mathrm{S}}$. 
\par
Hereafter, we consider Cooper pairs with following symmetries; 
even-$\omega$ spin-singlet even-parity, even-$\omega$ spin-triplet odd-parity, odd-$\omega$ spin-singlet odd-parity, and odd-$\omega$ spin-triplet 
even-parity symmetries for various  $h_s$. 
Spin-singlet pairing has a spin state 
$\mid\uparrow\downarrow\rangle-\mid\downarrow\uparrow\rangle$ 
(total pair spin $S_z=0$) while spin-triplet pairing has spin states 
$\mid\uparrow\uparrow\rangle$ ($S_z=1$), 
$\mid\downarrow\downarrow\rangle$ ($S_z=-1$), and 
$\mid\uparrow\downarrow\rangle+\mid\downarrow\uparrow\rangle$ ($S_z=0$). 
We present that odd-$\omega$ spin-triplet ($S_z=\pm 1$) 
$s$-wave pairing state exceeds conventional even-$\omega$ spin-singlet 
$d$-wave one. 
In the actual calculation, 
$\mu$ in Eq. (\ref{eq_green}) is determined to tune $0.8$ electrons 
per site. 
Temperature is chosen to be $T=0.01t$. 
$\omega_n$ has a value 
from $-(2n_{\mathrm{max}}-1)\pi T$ to $(2n_{\mathrm{max}}-1)\pi T$ with 
$n_{\mathrm{max}}=2048$. 
$\nu_m$ has a value from $-2m_{\mathrm{max}}\pi T$ 
to $2m_{\mathrm{max}}\pi T$ with $m_{\mathrm{max}}=2048$. 
We take $N=64\times 64$. 
A gap function is normalized with a condition 
$\sum_{k,\alpha,\beta}|\Delta_{\sigma\sigma'}^{\alpha\beta}(k)|^2=1$. 
We choose a value of $U$ for $f_{\mathrm{S}}$ 
to get 0.98 because $f_{\mathrm{S}}$ determines 
a scale of $\hat{V}_{\sigma\sigma'}(k,k')$, to which $\lambda$ is sensitive. 
\par
\begin{figure}[htbp]
\includegraphics[width=0.75\linewidth,keepaspectratio]
                  {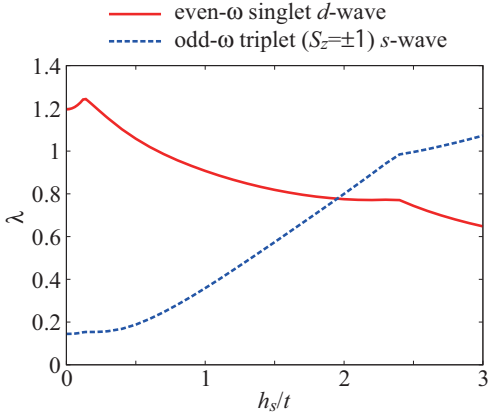}
 \caption{(Color online) $h_s$ dependences of $\lambda$ for leading pairing symmetries; even-$\omega$ spin-singlet $d$-wave and odd-$\omega$ spin-triplet ($S_z=\pm 1$) $s$-wave.}
\label{fig_h_lambda}
\end{figure}
\begin{figure}[htbp]
\includegraphics[width=0.8\linewidth,keepaspectratio]
                  {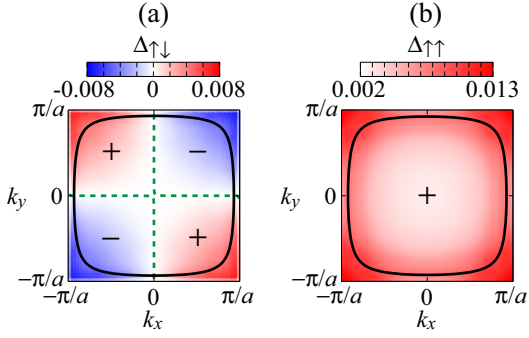}
 \caption{(Color online) ${\boldsymbol{k}}$ dependences of gap functions for (a) even-$\omega$ spin-singlet $d$-wave pairing at $h_s=0$ and (b) odd-$\omega$ spin-triplet ($S_z=1$) $s$-wave one at $h_s=2.6t$ in the band basis at $\omega_n=\pi T$. Dotted (green) and solid (black) lines denote nodal lines and the Fermi surfaces, respectively. ``$+$'' and ``$-$'' denote signs of gap functions.}
\label{fig_gap}
\end{figure}

%
%
Fig. \ref{fig_h_lambda} shows $h_s$ dependences of $\lambda$. 
Here, we plot only pairings which become dominant. 
Odd-$\omega$ spin-triplet ($S_z=\pm 1$) $s$-wave pairing state is 
the most stable one in a high $h_s$ region ($h_s\gtrsim 2t$) 
while even-$\omega$ spin-singlet $d$-wave pairing state is 
the most stable one in a low $h_s$ region ($0\leq h_s\lesssim 2t$). 
Figs. \ref{fig_gap} (a) and (b) show ${\boldsymbol{k}}$ dependences of 
gap functions for even-$\omega$ spin-singlet $d$-wave pairing at $h_s=0$ 
and odd-$\omega$ spin-triplet ($S_z=1$) $s$-wave one at $h_s=2.6t$, 
respectively, in the band basis. 
Here, we show only one of two bands where the Fermi surface exists 
for each pairing. 
The present odd-$\omega$ spin-triplet ($S_z=1$) $s$-wave pairing 
in a high $h_s$ region is on-site equal-spin one 
and has a weak modulation in real space. 
\par
\begin{figure}[htbp]
\includegraphics[width=0.75\linewidth,keepaspectratio]
                  {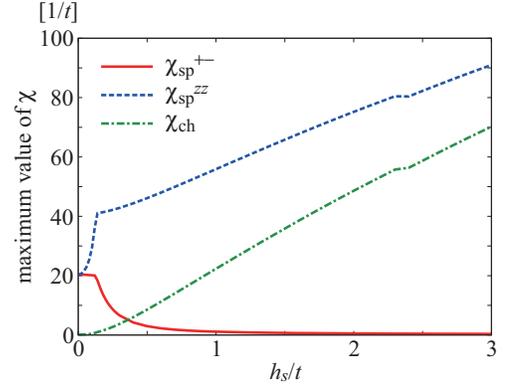}
 \caption{(Color online) $h_s$ dependences of $\hat{\chi}_{\mathrm{sp}}^{+-}(q)$, $\hat{\chi}_{\mathrm{sp}}^{zz}(q)$, and $\hat{\chi}_{\mathrm{ch}}(q)$. The maximum values after diagonalization are plotted.}
\label{fig_h_chisc}
\end{figure}

We discuss why the pairing competition between 
even-$\omega$ spin-singlet $d$-wave and 
odd-$\omega$ spin-triplet ($S_z=\pm 1$) $s$-wave pairings 
occurs under $h_s$ as shown in Fig. \ref{fig_h_lambda}. 
This pairing competition is caused by suppression of spin fluctuation 
in a $xy$ plane and enhancement of charge fluctuation 
originating from $h_s$. 
$\hat{\chi}_{\mathrm{sp}}^{+-}(q)$, $\hat{\chi}_{\mathrm{sp}}^{zz}(q)$, 
and $\hat{\chi}_{\mathrm{ch}}(q)$ depend on $h_s$ as shown in 
Fig. \ref{fig_h_chisc}. 
With increasing $h_s$, $\hat{\chi}_{\mathrm{sp}}^{+-}(q)$ decreases 
while $\hat{\chi}_{\mathrm{sp}}^{zz}(q)$ and 
$\hat{\chi}_{\mathrm{ch}}(q)$ increase. 
In other words, spin fluctuation in a $xy$ plane is suppressed while 
that in a $z$ direction and charge fluctuation are enhanced. 
In order to understand how these changes influence the 
symmetry of Cooper pair, 
we look at the single-band Hubbard model without $h_s$. 
In the case where spin or charge fluctuation dominates, 
an effective pairing interaction $V_{\sigma\sigma'}(k,k')$ 
in the single-band Hubbard model without $h_s$ is written as 
\begin{align}
&V_{\sigma\bar{\sigma}}(k,k') \notag \\
&\mspace{10mu}\sim\mspace{13mu}\frac{1}{2}U^2\left[2\chi_{\mathrm{sp}}^{+-}(k+k')+\chi_{\mathrm{sp}}^{zz}(k-k')-\chi_{\mathrm{ch}}(k-k')\right], 
\label{eq_single_band_interaction_ap}
\\
&V_{\sigma\sigma}(k,k') \notag \\
&\mspace{10mu}\sim-\frac{1}{2}U^2\left[\mspace{126mu}\chi_{\mathrm{sp}}^{zz}(k-k')+\chi_{\mathrm{ch}}(k-k')\right], 
\label{eq_single_band_interaction_p}
\end{align}
with the transverse spin susceptibility $\chi_{\mathrm{sp}}^{+-}(q)$, 
the longitudinal spin susceptibility $\chi_{\mathrm{sp}}^{zz}(q)$, and 
the charge susceptibility $\chi_{\mathrm{ch}}(q)$, 
which are positive numbers. 
As is shown in Eqs. (\ref{eq_single_band_interaction_ap}) and 
(\ref{eq_single_band_interaction_p}), suppression of the spin fluctuation 
in a $xy$ plane destabilizes pairing with $S_z=0$ while 
enhancement of charge fluctuation destabilizes (stabilizes) pairing 
with $S_z=0$ ($S_z=\pm 1$). 
By these features, $h_s$ causes the pairing competition 
between even-$\omega$ spin-singlet $d$-wave and 
odd-$\omega$ spin-triplet ($S_z=\pm 1$) $s$-wave pairings. 
\par
\begin{figure}[htbp]
\includegraphics[width=0.99\linewidth,keepaspectratio]
                  {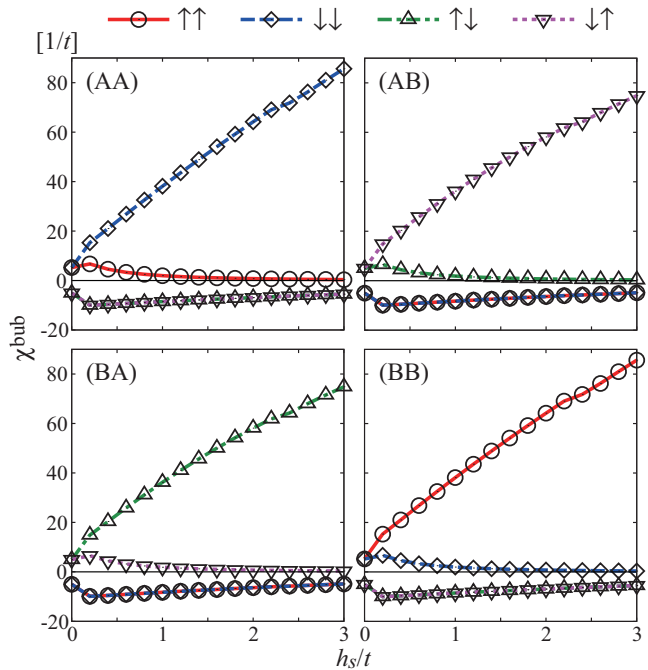}
 \caption{(Color online) $h_s$ dependences of (AA) $\chi_{\sigma\sigma'}^{\mathrm{bub,AA}}(q)$, (AB) $\chi_{\sigma\sigma'}^{\mathrm{bub,AB}}(q)$, (BA) $\chi_{\sigma\sigma'}^{\mathrm{bub,BA}}(q)$, and (BB) $\chi_{\sigma\sigma'}^{\mathrm{bub,BB}}(q)$. In each panel, the values at $q$-point where the absolute values get maximum are plotted for all four combinations of spins.}
\label{fig_h_chib}
\end{figure}

Here, you may wonder why not even-$\omega$ spin-triplet ($S_z=\pm 1$) 
$p$-wave but odd-$\omega$ spin-triplet ($S_z=\pm 1$) $s$-wave pairing 
is stabilized. 
The reason is that an electron with each spin feels inhomogeneity 
in real space due to the background AFO. 
%
Even-$\omega$ spin-triplet ($S_z=\pm 1$) $p$-wave pairing tends to be 
equal-spin pairing between the nearest sites. 
However, in a high $h_s$ region, 
$\uparrow$-electron-rich and 
$\downarrow$-electron-rich sites are adjacent. 
Therefore, 
$\chi_{\sigma\sigma}^{\mathrm{bub},\alpha\bar{\alpha}}(q)$, which mediates 
equal-spin pairing between the nearest sites, is suppressed 
while $\chi_{\downarrow\downarrow}^{\mathrm{bub,AA}}(q)$ and 
$\chi_{\uparrow\uparrow}^{\mathrm{bub,BB}}(q)$, which mediate 
on-site equal-spin pairing, are enhanced 
(see Fig. \ref{fig_h_chib}).  
Thus even-$\omega$ spin-triplet ($S_z=\pm 1$) $p$-wave pairing 
is not allowed there and odd-$\omega$ spin-triplet ($S_z=\pm 1$) 
$s$-wave one appears. 
\par
Next, we discuss why $\hat{\chi}_{\mathrm{sp}}^{+-}(q)$ decreases 
while $\hat{\chi}_{\mathrm{sp}}^{zz}(q)$ and 
$\hat{\chi}_{\mathrm{ch}}(q)$ increase with increasing $h_s$ 
as shown in Fig. \ref{fig_h_chisc}. 
The decrease of $\hat{\chi}_{\mathrm{sp}}^{+-}(q)$ and the increase of 
$\hat{\chi}_{\mathrm{sp}}^{zz}(q)$ can be interpreted as 
the suppression of spin fluctuation in a $xy$ plane and the enhancement of 
that in a $z$ direction, respectively, due to 
$h_s$ along a $z$ direction. 
The increase of $\hat{\chi}_{\mathrm{ch}}(q)$ is caused by inhomogeneity 
in real space for an electron with each spin due to $h_s$. 
$\hat{\chi}_{\sigma\sigma'}^{\mathrm{bub}}(q)$, which gives 
$\hat{\chi}_{\mathrm{sp}}^{zz}(q)$ and $\hat{\chi}_{\mathrm{ch}}(q)$ 
as written in Eqs. (\ref{eq_chi_sp_zz}) and (\ref{eq_chi_ch}), 
depends on $h_s$ as shown in Fig. \ref{fig_h_chib}. 
$\chi_{\sigma\sigma}^{{\mathrm{bub}},\alpha\beta}(q)$ and 
$\chi_{\sigma\bar{\sigma}}^{{\mathrm{bub}},\alpha\beta}(q)$ 
are opposite in sign to each other. 
In a low $h_s$ region, 
they have comparable absolute values and 
cancel out each other in the summation of 
right side of 
$\chi_{\mathrm{ch}}^{\alpha\beta}(q)$ in Eq. (\ref{eq_chi_ch}), 
while they enhance each other in that of right side of 
$\chi_{\mathrm{sp}}^{zz,\alpha\beta}(q)$ 
in Eq. (\ref{eq_chi_sp_zz}). 
%
%
%
This picture is similar to that in the single-band Hubbard model 
without $h_s$. 
In a high $h_s$ region, on the other hand, the above 
picture breaks down since 
only the susceptibility 
between a specific combination of 
spins is prominently enhanced. 
As shown in Fig. \ref{fig_h_chib}, the magnitudes of 
$\chi_{\downarrow\downarrow}^{\mathrm{bub,AA}}(q)$ in a panel (AA), 
$\chi_{\downarrow\uparrow}^{\mathrm{bub,AB}}(q)$ in a panel (AB), 
$\chi_{\uparrow\downarrow}^{\mathrm{bub,BA}}(q)$ in a panel (BA), and
$\chi_{\uparrow\uparrow}^{\mathrm{bub,BB}}(q)$ in a panel (BB) 
are prominently enhanced. 
This feature arises from the inhomogeneity in real space for an electron 
with each spin, 
$i.e.$, $\downarrow$-electron-richness on a sublattice A and 
$\uparrow$-electron-richness on a sublattice B due to the presence of $h_s$. 
In the present situation, the above cancellation for low $h_s$ 
in the expression of 
$\hat{\chi}_{\mathrm{ch}}(q)$ does not work any more. 
Thus, in a high $h_s$ region, $\hat{\chi}_{\mathrm{ch}}(q)$ becomes 
dominant in addition to $\hat{\chi}_{\mathrm{sp}}^{zz}(q)$. 
\par
%
%
In summary, we have studied  pairing symmetry in the superconducting state 
coexisting with AF by solving the linearized 
$\mathrm{\acute{E}}$liashberg's equation 
in the Hubbard model with a staggered field using the RPA. 
As a result, we have found that odd-$\omega$ equal-spin-triplet 
$s$-wave pairing can be realized.  
This is caused by the suppression 
of spin fluctuation in a plane perpendicular to 
an antiferromagnetic moment 
and the enhancement of charge fluctuation due to the background AFO. 
In particular, the enhancement of charge fluctuation originates from 
inhomogeneity in real space for an electron with each spin induced by 
the background AFO. 
%
The present result contradicts a naive picture 
that AFO might help antiferromagnetic 
spin fluctuation and favor even-$\omega$ spin-singlet $d$-wave pairing. 
We hope that the odd-$\omega$ spin-triplet pairing will be 
verified experimentally in strongly correlated systems 
where SC coexists with AF. 
\par
%
%
K.S. acknowledges support by JSPS. 
%
%


\begin{thebibliography}{99}

\bibitem{Ginzburg}
V. L. Ginzburg, 
Sov. Phys. JETP {\bf 4}, 153 (1957); 
P. W. Anderson and H. Suhl, 
Phys. Rev. {\bf 116}, 898 (1959). 

\bibitem{UGe2}
S. S. Saxena, P. Agrwal, A. Ahilan, F. M. Grosche, R. K. W. Haselwimmer, 
M. J. Steiner, E. Pugh, I. R. Walker, S. R. Julian, P. Monthoux, 
G. G. Lonzarich, A. Huxley, I. Sheiken, D. Braithwaite, and J. Flouquet, 
Nature {\bf 406}, 587 (2000). 

\bibitem{URhGe}
D. Aoki, A. Huxley, E. Ressouche, D. Braithwaite, J. Flouquet, 
J.-P. Brison, E. Lhotel, and C. Paulsen, 
Nature {\bf 413}, 613 (2001). 

\bibitem{CeCu2Si2}
Y. Kitaoka, K. Ishida, Y. Kawasaki, O. Trovarelli, C. Geibel, 
and F. Steglich, 
J. Phys.: Condens. Matter {\bf 13}, L79 (2001). 

\bibitem{CeRhIn5}
T. Mito, S. Kawasaki, Y. Kawasaki, G.-q. Zheng, Y. Kitaoka, D. Aoki, 
Y. Haga, and Y. $\mathrm{\bar{O}}$nuki, 
Phys. Rev. Lett. {\bf 90}, 077004 (2003). 

\bibitem{CeCu2Si2_NQR}
Y. Kawasaki, K. Ishida, K. Obinata, K. Tabuchi, K. Kashima, and Y. Kitaoka, 
Phys. Rev. B {\bf 66}, 224502 (2002). 

\bibitem{CeRhIn5_NQR}
S. Kawasaki, T. Mito, Y. Kawasaki, G.-q. Zheng, Y. Kitaoka, D. Aoki, 
Y. Haga, and Y. $\mathrm{\bar{O}}$nuki, 
Phys. Rev. Lett. {\bf 91}, 137001 (2003). 

\bibitem{CeRhIn5_specific_heat}
R. A. Fisher, F. Bouquet, N. E. Phillips, M. F. Hundley, P. G. Pagliuso, 
J. L. Sarrao, Z. Fisk, and J. D. Thompson, 
Phys. Rev. B {\bf 65}, 224509 (2002). 

\bibitem{Berezinskii}
V. L. Berezinskii, 
JETP Lett. {\bf 20}, 287 (1974).

\bibitem{Fuseya}
Y. Fuseya, H. Kohno, and K. Miyake, 
J. Phys. Soc. Jpn. {\bf 72}, 2914 (2003). 

\bibitem{odd1}
A. Balatsky and E. Abrahams, 
Phys. Rev. B {\bf 45}, 13125 (1992). 

\bibitem{odd2}
P. Coleman, A. Georges, and A. M. Tsvelik, 
J. Phys.: Condens. Matter {\bf 9}, 345 (1997). 

\bibitem{odd3}
M. Vojta and E. Dagotto, 
Phys. Rev. B {\bf 59}, R713 (1999). 

\bibitem{odd4}
K. Shigeta, S. Onari, K. Yada, and Y. Tanaka, 
Phys. Rev. B {\bf 79}, 174507 (2009). 

\bibitem{Hotta}
T. Hotta, 
J. Phys. Soc. Jpn. {\bf 78}, 123710 (2009). 

\bibitem{Solenov}
D. Solenov, I. Martin, and D. Mozyrsky, 
Phys. Rev. B {\bf 79}, 132502 (2009). 

\bibitem{Kusunose}
H. Kusunose, Y. Fuseya, and K. Miyake, 
arXiv:1011.4712; 
H. Kusunose, Y. Fuseya, and K. Miyake, 
arXiv:1012.5333. 

\bibitem{odd5}
F. S. Bergeret, A. F. Volkov, and K. B. Efetov, 
Phys. Rev. Lett. {\bf 86}, 4096 (2001). 

\bibitem{odd6}
Y. Tanaka and A. A. Golubov, 
Phys. Rev. Lett. {\bf 98}, 037003 (2007). 

\bibitem{odd7}
Y. Tanaka, A. A. Golubov, S. Kashiwaya, and M. Ueda, 
Phys. Rev. Lett. {\bf 99}, 037005 (2007). 


\end{thebibliography}
\end{document}